# Optimal Energy Management Policies for Energy Harvesting Sensor Nodes

Vinod Sharma, *Senior Member IEEE*, Utpal Mukherji, *Senior Member IEEE*, Vinay Joseph and Shrey Gupta

### Abstract

We study a sensor node with an energy harvesting source. The generated energy can be stored in a buffer. The sensor node periodically senses a random field and generates a packet. These packets are stored in a queue and transmitted using the energy available at that time. We obtain energy management policies that are throughput optimal, i.e., the data queue stays stable for the largest possible data rate. Next we obtain energy management policies which minimize the mean delay in the queue. We also compare performance of several easily implementable sub-optimal energy management policies. A greedy policy is identified which, in low SNR regime, is throughput optimal and also minimizes mean delay.

## I. INTRODUCTION

Sensor networks consist of a large number of small, inexpensive sensor nodes. These nodes have small batteries with limited power and also have limited computational power and storage space. When the battery of a node is exhausted, it is not replaced and the node dies. When sufficient number of nodes die, the network may not be able to perform its designated task. Thus the life time of a network is an important characteristic of a sensor network ([4]) and it is tied up with the life time of a node.

Various studies have been conducted to increase the life time of the battery of a node by reducing the energy intensive tasks, e.g., reducing the number of bits to transmit ([22], [5]), making a node to go

Vinod Sharma, Utpal Mukherji, Vinay Joseph are with the Dept of Electrical Communication Engineering, IISc, Bangalore, India. Email: { vinod,utpal,vinay }@ece.iisc.ernet.in

Shrey Gupta is with the Dept of Computer Science and Engineering, Indian Institute of Technology, Guwahati, India. Email : shrey@iitg.ernet.in

into power saving modes: (sleep/listen) periodically ([28]), using energy efficient routing ([30], [25]) and MAC ([31]). Studies that estimate the life time of a sensor network include [25]. A general survey on sensor networks is [1] which provides many more references on these issues.

In this paper we focus on increasing the life time of the battery itself by energy harvesting techniques ([14], [21]). Common energy harvesting devices are solar cells, wind turbines and piezo-electric cells, which extract energy from the environment. Among these, solar harvesting energy through photo-voltaic effect seems to have emerged as a technology of choice for many sensor nodes ([21], [23]). Unlike for a battery operated sensor node, now there is potentially an *infinite* amount of energy available to the node. Hence energy conservation need not be the dominant theme. Rather, the issues involved in a node with an energy harvesting source can be quite different. The source of energy and the energy harvesting device may be such that the energy cannot be generated at all times (e.g., a solar cell). However one may want to use the sensor nodes at such times also. Furthermore the rate of generation of energy can be limited. Thus one may want to match the energy generation profile of the harvesting source with the energy consumption profile of the sensor node. If the energy can be *stored* in the sensor node then this matching can be considerably simplified. But the energy storage device may have limited capacity. Thus, one may also need to modify the energy consumption profile of the sensor node so as to achieve the desired objectives with the given energy harvesting source. It should be done in such a way that the node can perform satisfactorily for a long time, i.e., energy starvation at least, should not be the reason for the node to die. In [14] such an energy/power management scheme is called *energy neutral operation* (if the energy harvesting source is the only energy source at the node, e.g., the node has no battery). Also, in a sensor network, the routing and relaying of data through the network may need to be suitably modified to match the energy generation profiles of different nodes, which may vary with the nodes.

In the following we survey the literature on sensor networks with energy harvesting nodes. Early papers on energy harvesting in sensor networks are [15] and [24]. A practical solar energy harvesting sensor node prototype is described in [12]. A good recent contribution is [14]. It provides various deterministic theoretical models for energy generation and energy consumption profiles (based on $(\sigma, \rho)$ traffic models in [8]) and provides conditions for energy neutral operation. In [11] a sensor node is considered which is sensing certain interesting events. The authors study optimal sleep-wake cycles such that event detection probability is maximized. This problem is also studied in [3]. A recent survey is [21] which also provides an optimal sleep-wake cycle for solar cells so as to obtain QoS for a sensor node.

In this paper we study a sensor node with an energy harvesting source. The motivating application is estimation of a random field which is one of the canonical applications of sensor networks. The above mentioned theoretical studies are motivated by other applications of sensor networks. In our application, the sensor nodes sense the random field periodically. After sensing, a node generates a packet (possibly after efficient compression). This packet needs to be transmitted to a central node, possibly via other sensor nodes. In an energy harvesting node, sometimes there may not be sufficient energy to transmit the generated packets (or even sense) at regular intervals and then the node may need to store the packets till they are transmitted. The energy generated can be stored (possibly in a finite storage) for later use.

Initially we will assume that most of the energy is consumed in transmission only. We will relax this assumption later on. We find conditions for energy neutral operation of the system, i.e., when the system can work forever and the data queue is stable. We will obtain policies which can support maximum possible data rate.

We also obtain energy management (power control) policies for transmission which minimize the mean delay of the packets in the queue.

Our energy management policies can be used with sleep-wake cycles. Our policies can be used on a faster time scale during the wake period of a sleep-wake cycle. When the energy harvesting profile generates minimal energy (e.g., in solar cells) then one may schedule the sleep period.

We have used the above energy mangement policies at a MAC (Multiple Access Channel) used by energy harvesting sensor nodes in [27].

We are currently investigating appropriate routing algorithms for a network of energy harvesting sensor nodes.

The paper is organized as follows. Section II describes the model and provides the assumptions made for data and energy generation. Section III provides conditions for energy neutral operation. We obtain stable, power control policies which are throughput optimal. Section IV obtains the power control policies which minimize the mean delay via Markov decision theory. A greedy policy is shown to be throughput optimal and provides minimum mean delays for linear transmission. Section V provides a throughput optimal policy when the energy consumed in sensing and processing is nonnegligible. A sensor node with a fading channel is also considered. Section VI provides simulation results to confirm our theoretical findings and compares various energy management policies. Section VII concludes the paper. The appendix provides proof of the lemma used in proving existence of an optimal policy.

## II. Model and notation

In this section we present our model for a single energy harvesting sensor node.

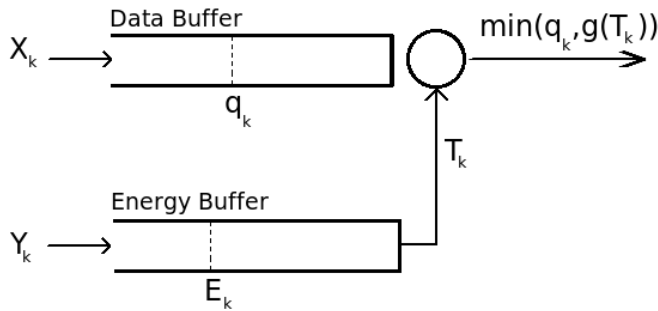

Fig. 1.  The model

We consider a sensor node (Fig. 1) which is sensing a random field and generating packets to be transmitted to a central node via a network of sensor nodes. The system is slotted. During slot $k$ (defined as time interval $[k, k+1]$, i.e., a slot is a unit of time) $X_k$ bits are generated by the sensor node. Although the sensor node may generate data as packets, we will allow arbitrary fragmentation of packets during transmission. Thus, packet boundaries are not important and we consider bit strings (or just fluid). The bits $X_k$ are eligible for transmission in $(k + 1)$st slot. The queue length (in bits) at time $k$ is $q_k$. The sensor node is able to transmit $g(T_k)$ bits in slot $k$ if it uses energy $T_k$. We assume that transmission consumes most of the energy in a sensor node and ignore other causes of energy consumption (this is true for many low quality, low rate sensor nodes ([23])). This assumption will be removed in Section V. We denote by $E_k$ the energy available in the node at time $k$. The sensor node is able to replenish energy by $Y_k$ in slot $k$.

We will initially assume that $\{X_k\}$ and $\{Y_k\}$ are *iid* but will generalize this assumption later. It is important to generalize this assumption to capture realistic traffic streams and energy generation profiles.

The processes $\{q_k\}$ and $\{E_k\}$ satisfy

$$q_{k+1} = (q_k - g(T_k))^+ + X_k, \tag{1}$$

$$E_{k+1} = (E_k - T_k) + Y_k. \tag{2}$$

where $T_k \leq E_k$. This assumes that the data buffer and the energy storage buffer are infinite. If in practice these buffers are large enough, this is a good approximation. If not, even then these results provide important insights and the policies obtained often provide good performance for the finite buffer case.

The function $g$ will be assumed to be monotonically non-decreasing. An important such function is given by Shannon's capacity formula

$$g(T_k) = \frac{1}{2} log(1 + \beta T_k)$$

for Gaussian channels where $\beta$ is a constant such that $\beta \, T_k$ is the SNR. This is a non-decreasing concave function. At low values of $T_k$, $g(T_k) \sim \beta_1 \, T_k$, i.e., $g$ becomes a linear function. Since sensor nodes are energy constrained, this is a practically important case. Thus in the following we limit our attention to linear and concave nondecreasing functions $g$. We will also assume that $g(0) = 0$ which always holds in practice.

Many of our results (especially the stability results) will be valid when $\{X_k\}$ and $\{Y_k\}$ are stationary, ergodic. These assumptions are general enough to cover most of the stochastic models developed for traffic (e.g., Markov modulated) and energy harvesting.

Of course, in practice, statistics of the traffic and energy harvesting models will be time varying (e.g., solar cell energy harvesting will depend on the time of day). But often they can be approximated by piecewise stationary processes. For example, energy harvesting by solar cells could be taken as being stationary over one hour periods. Then our results could be used over these time periods. Often these periods are long enough for the system to attain (approximate) stationarity and for our results to remain meaningful.

In Section III we study the stability of this queue and identify easily implementable energy management policies which provide good performance.

## III. STABILITY

We will obtain a necessary condition for stability. Then we present a transmission policy which achieves the necessary condition, i.e., the policy is throughput optimal. The mean delay for this policy is not minimal. Thus, we obtain other policies which provide lower mean delay. In the next section we will consider optimal policies.

Let us assume that we have obtained an (asymptotically) stationary and ergodic transmission policy $\{T_k\}$ which makes $\{q_k\}$ (asymptotically) stationary with the limiting distribution independent of $q_0$. Taking $\{T_k\}$ asymptotically stationary seems to be a natural requirement to obtain (asymptotic) stationarity of $\{q_k\}$.

**Lemma 1** Let $g$ be concave nondeceasing and $\{X_k\}, \{Y_k\}$ be stationary, ergodic sequences. For $\{T_k\}$ to be an asymptotically stationary, ergodic energy management policy that makes $\{q_k\}$ asymptotically stationary with a proper stationary distribution $\pi$ it is necessary that $E[X_k] < E_\pi[g(T)] \le g(E[Y])$.

**Proof:** Let the system start with $q_0 = E_0 = 0$. Then for each $n$, $n^{-1}\sum_{k=1}^{n} T_k \le n^{-1}\sum_{k=1}^{n} Y_k + \frac{Y_0}{n}$. Thus, if $n^{-1}\sum_{k=1}^{n} T_k \to E[T]$ a.s., then $E[T] \le E[Y]$. Also then $n^{-1}\sum_{k=1}^{n} g(T_k) \to E[g(T)]$ a.s.

Thus from results on G/G/1 queues [6], $E[g(T)] > E[X]$ is needed for the (asymptotic) stationarity of $\{q_k\}$. If $g$ is linear then the above inequalities imply that for stationarity of $\{q_k\}$ we need

$$
\begin{aligned}
E[X] &< E[g(T)] = g(E[T]) \\
&\le g(E[Y]) = E[g(Y)].
\end{aligned}
\tag{3}
$$

If $g$ is concave, then we need

$$
E[X] < E[g(T)] \le g(E[T]) \le g(E[Y]).
\tag{4}
$$

Thus $E[X] < g(E[Y])$ is a necessary condition to get an (asymptotically) stationary sequence $\{g(T_k)\}$ which provides an asymptotically stationary $\{q_k\}$. ∎

Let

$$
T_k = min(E_k, E[Y] - \epsilon)
\tag{5}
$$

where $\epsilon$ is an appropriately chosen small constant (see statement of Theorem 1). We show that it is a throughput optimal policy, i.e., using this $T_k$ with $g$ satisfying the assumptions in Lemma 1, $\{q_k\}$ is asymptotically stationary and ergodic.

**Theorem 1** If $\{X_k\}, \{Y_k\}$ are stationary, ergodic, $g$ is continuous, nondecreasing, concave then if $E[X_k] < g(E[Y])$, (5) makes the queue stable (with $\epsilon > 0$ such that $E[X] < g(E[Y] - \epsilon)$), i.e., it has a unique, stationary, ergodic distribution and starting from any initial distribution, $q_k$ converges in total variation to the stationary distribution.

**Proof:** If we take $T_k = min(E_k, E[Y] - \epsilon)$ for any arbitrarily small $\epsilon > 0$, then from (2), $E_k \nearrow \infty$ a.s. and $T_k \nearrow E[Y] - \epsilon$. a.s. If $g$ is continuous in a neighbourhood of $E[Y]$ then by monotonicity of $g$ we also get $g(T_k) \nearrow g(E[Y] - \epsilon)$ a.s. Hence $E[g(T_k)] \nearrow g(E[Y] - \epsilon)$. We also get $E[T_k] \nearrow E[Y] - \epsilon$. Thus $\{g(T_k)\}$ is asymptotically stationary and ergodic. Therefore, from G/G/1 queue results [6], [19] for $T_k = min(E_k, E[Y] - \epsilon)$, $E[X] < g(E[Y] - \epsilon)$ is a sufficient condition for $\{q_k\}$ to be asymptotically stationary and ergodic whenever $\{X_k\}$ is stationary and ergodic. The other conclusions also follow. Since $g$ is non-decreasing and $g(0) = 0$, $E[X_k] < g(E[Y])$ implies that there is an $\epsilon > 0$ such that $E[X] < g(E[Y] - \epsilon)$. ∎

Henceforth we denote the policy (5) by TO.

From results on GI/GI/1 queues ([2]), if $\{X_k\}$ are *iid*, $E[X] < g(E[Y]), T_k = min(E_k, E[Y] - \epsilon)$ and $E[X^\alpha] < \infty$ for some $\alpha > 1$ then the stationary solution $\{q_k\}$ of (1) satisfies $E[q^{\alpha-1}] < \infty$.

Taking $T_k = Y_k$ for all $k$ will provide stability of the queue if $E[X] < E[g(Y)]$. If $g$ is linear then this coincides with the necessary condition. If $g$ is strictly concave then $E[g(Y)] < g(E[Y])$ unless $Y \equiv E[Y]$. Thus $T_k = Y_k$ provides a strictly smaller stability region. We will be forced to use this policy if there is no buffer to store the energy harvested. This shows that storing energy allows us to have a larger stability region. We will see in Section VI that storing energy can also provide lower mean delays.

Although TO is a throughput optimal policy, if $q_k$ is small, we may be wasting some energy. Thus, it appears that this policy does not minimize mean delay. It is useful to look for policies which minimize mean delay. Based on our experience in [26], the Greedy policy

$$T_k = min(E_k, f(q_k)) \qquad (6)$$

where $f = g^{-1}$, looks promising. In Theorem 2, we will show that the stability condition for this policy is $E[X] < E[g(Y)]$ which is optimal for linear $g$ but strictly suboptimal for a strictly concave $g$. We will also show in Section IV that when $g$ is linear, (6) is not only throughput optimal, it also minimizes long term mean delay.

For concave $g$, we will show via simulations that (6) provides less mean delay than TO at low load. However since its stability region is smaller than that of the TO policy, at $E[X]$ close to $E[g(Y)]$, the Greedy performance rapidly deteriorates. Thus it is worthwhile to look for some other good policy. Notice that the TO policy wastes energy if $q_k < g(E[Y] - \epsilon)$. Thus we can improve upon it by saving the energy $(E[Y] - \epsilon - g^{-1}(q_k))$ and using it when the $q_k$ is greater than $g(E[Y] - \epsilon)$. However for $g$ a log function, using a large amount of energy $t$ is also wasteful even when $q_k > g(t)$. Taking into account these facts we improve over the TO policy as

$$T_k = min(g^{-1}(q_k), E_k, 0.99(E[Y] + 0.001(E_k - cq_k)^+)) \qquad (7)$$

where $c$ is a positive constant. The improvement over the TO also comes from the fact that if $E_k$ is large,

we allow $T_k > E[Y]$ but only if $q_k$ is not very large. The constants 0.99 and 0.001 were chosen by trial and error from simulations after experimenting with different scenarios. We will see in Section VI via simulations that the policy, to be denoted by MTO can indeed provide lower mean delays than TO at loads above $E[g(Y)]$.

One advantage of (5) over (6) and (7) is that while using (5), after some time $T_k = E[Y] - \epsilon$. Also, at any time, either one uses up all the energy or uses $E[Y] - \epsilon$. Thus one can use this policy even if exact information about $E_k$ is not available (measuring $E_k$ may be difficult in practice). In fact, (5) does not need even $q_k$ while (6) either uses up all the energy or uses $f(q_k)$ and hence needs only $q_k$ exactly.

Now we show that under the greedy policy (6) the queueing process is stable when $E[X] < E[g(Y)]$. In next few results we assume that the energy buffer is finite, although large. For this case Lemma 1 and Theorem 1 also hold under the same assumptions with slight modifications in their proofs.

**Theorem 2** If the energy buffer is finite, i.e., $E_k \leq \bar{e} < \infty$ (but $\bar{e}$ is large enough) and $E[X] < E[g(Y)]$ then under the greedy policy (6), $(q_k, E_k)$ has an Ergodic set.

**Proof:** To prove that $(q_k, E_k)$ has an ergodic set [20], we use the Lyapunov function $h(q, e) = q$ and show that this has a negative drift outside a large enough set of state space

$$A \overset{\triangle}{=} \{(q, e) : q + e > \beta\}$$

where $\beta > 0$ is appropriately chosen. If we take $\beta$ large enough, because $e \leq \bar{e}$, $(q, e) \in A$ will ensure that $q$ is appropriately large. We will specify our requirements on this later.

For $(q, e) \in A, M > 0$ fixed, since we are using greedy policy

$$
\begin{aligned}
& E[h(q_{k+M}, E_{k+M}) - h(q_k, E_k)|(q_k, E_k) = (q, e)] \\
= \ & E[(q - g(T_k) + X_k - g(T_{k+1})) + X_{k+1} - \ldots \\
& \ldots - g(T_{k+M-1}) + X_{k+M-1} - q|(q_k, E_k) = (q, e)].
\end{aligned}
\tag{8}
$$

Because $T_n \leq E_n \leq \bar{e}$, we can take $\beta$ large enough such that the RHS of (8) equals

$$E[q + \sum_{n=k}^{k+M-1} X_n - \sum_{n=k+1}^{k+M-1} g(T_n) - g(e) - q|(q_k, E_k) = (q, e)].$$

Thus to have (8) less than $-\epsilon_2$ for some $\epsilon_2 > 0$, it is sufficient that

$$ME[X] < E\left[\sum_{n=k+1}^{k+M-1} g(T_n)\right] + g(e).$$

This can be ensured for any $e$ because we can always take $T_n \geq min(\bar{e}, Y_{n-1})$ with probability $> 1 - \delta$ (for any given $\delta > 0$) for $n = k+1, \ldots, k+M-1$ if in addition we also have $ME[X] < (M-1)E[g(Y)]$ and $\bar{e}$ is large enough. This can be ensured for a large enough $M$ because $E[X] < E[g(Y)]$. ∎

The above result will ensure that the Markov chain $\{(q_k, E_k)\}$ is ergodic and hence has a unique stationary distribution if $\{(q_k, E_k)\}$ is irreducible. A sufficient condition for this is $0 < P[X_k = 0] < 1$ and $0 < P[Y_k = 0] < 1$ because then the state $(0,0)$ can be reached from any state with a positive probability. In general, $\{(q_k, E_k)\}$ can have multiple ergodic sets. Then, depending on the initial state, $\{(q_k, E_k)\}$ will converge to one of the ergodic sets and the limiting distribution depends on the initial conditions.

## IV. Optimal Policies

In this section we choose $T_k$ at time $k$ as a function of $q_k$ and $E_k$ such that

$$E\left[\sum_{k=0}^{\infty} \alpha^k \; q_k\right]$$

is minimized where $0 < \alpha < 1$ is a suitable constant. The minimizing policy is called $\alpha$-discount optimal. When $\alpha = 1$, we minimize

$$\limsup_{n \to \infty} \frac{1}{n} E\left[\sum_{k=0}^{n-1} q_k\right].$$

This optimizing policy is called average cost optimal. By Little's law [2] an average cost optimal policy also minimizes mean delay. If for a given $(q_k, e_k)$, the optimal policy $T_k$ does not depend on the past values, and is time invariant, it is called a stationary Markov policy.

If $\{X_k\}$ and $\{Y_k\}$ are Markov chains then these optimization problems are Markov Decision Problems (MDP). For simplicity, in the following we consider these problems when $\{X_k\}$ and $\{Y_k\}$ are *iid*. We obtain the existence of optimal $\alpha$-discount and average cost stationary Markov policies.

**Theorem 3** If $g$ is continuous and the energy buffer is finite, i.e., $e_k \leq \bar{e} < \infty$ then there exists an optimal $\alpha$-discounted Markov stationary policy. If in addition $E[X] < g(E[Y])$ and $E[X^2] < \infty$, then there exists an average cost optimal stationary Markov policy. The optimal cost $v$ does not depend on the initial state. Also, then the optimal $\alpha$-discount policies tend to an optimal average cost policy as $\alpha \to 1$. Furthermore, if $v_\alpha(q,e)$ is the optimal $\alpha$-discount cost for the initial state $(q,e)$ then

$$\lim_{\alpha \to 1}(1-\alpha)\,\inf_{(q,e)}v_\alpha(q,e) = v$$

**Proof:** We use Prop. 2.1 in [29] to obtain the existence of an optimal $\alpha$-discount stationary Markov policy. For this it is sufficient to verify the condition $(W)$ in [29]. The actions possible in state $(q_k, E_k) = (q,e)$ are $0 \le T_k \le e$. This forms a compact subset of the action space. Also this mapping is upper and lower semicontinuous. Under action $t$, the next state becomes $((q - g(t))^+ + X_k, e - t + Y_k)$. When $g$ is continuous, the mapping $t \mapsto ((q - g(t))^+ + X_k, e - t + Y_k)$ is a.s. continuous and hence the transition probability is continuous under weak convergence topology. In fact it converges under the stronger topology of setwise convergence. Also, the cost $(q,e) \mapsto q$ is continuous. Thus condition $(W)$ in [29] is satisfied. Not only we get existence of $\alpha$-discount optimal policy, from [10], we also get $v_n(q,e) \to v(q,e)$ as $n \to \infty$ where $v_n(q,e)$ is $n$-step optimal $\alpha$-discount cost.

To get the existence of an average cost optimal stationary Markov policy, we use Theorem 3.8 in [29]. This requires satisfying condition $(B)$ in [29] in addition to condition $(W)$. Let $J_\alpha(\delta, (q,e))$ be the $\alpha$-discount cost under policy $\delta$ with initial state $(q,e)$. Also let

$$m_\alpha = \inf_{(q,e)}v_\alpha(q,e).$$

Then we need to show that

$$sup_{\alpha<1}(v_\alpha(q,e) - m_\alpha) < \infty \tag{9}$$

for all $(q,e)$.

For this we use the TO policy described in Section II. We have shown that for this policy there is a unique stationary distribution and if $E[X^2] < \infty$ then $E[q] < \infty$ under stationarity.

Next we use the facts that $v_\alpha(q,e)$ is non decreasing in $q$ and non increasing in $e$. We will prove these at the end of this proof. Then $m_\alpha = v_\alpha(0, \bar{e})$.

Let $\tau$ be the first time $q_k = 0, E_k = \bar{e}$ when we use the TO policy. Under our conditions $E[\tau] < \infty$ if $q_0 = 0$ for any $e_0 = e$. Also, then

$$v_\alpha(q,e) \le E\left[\sum_{k=0}^{\tau-1}\alpha^k q_k | q_0 = q, e_0 = e\right] + v_\alpha(0, \bar{e}).$$

Thus,

$$v_\alpha(q,e) - v_\alpha(0,\bar{e}) \leq E\left[\sum_{k=0}^{\tau-1} q_k | q_0 = q, e_0 = 0\right].$$

For notational convenience in the following inequalities we omit writing conditioning on $q_0 = q, e_0 = e$. The RHS

$$\begin{aligned}
&\leq E\left[\sum_{k=0}^{\tau-1}(q + \sum_{l=0}^{k} X_l)\right] \\
&\leq E\left[\sum_{k=0}^{\tau-1}(q + \sum_{l=0}^{\tau-1} X_l)\right] \\
&= qE[\tau] + E\left[\tau \sum_{l=0}^{\tau-1} X_l\right] \\
&\leq qE[\tau] + E[\tau^2]^{\frac{1}{2}} E\left[(\sum_{l=0}^{\tau-1} X_l)^2\right]^{\frac{1}{2}}.
\end{aligned}$$

Since $[X_k]$ are iid and $\tau$ is a stopping time, $E\left[(\sum_{l=0}^{\tau-1} X_l)^2\right] < \infty$ if $E[\tau^2] < \infty$ and $E[X^2] < \infty$ ([9]). In Lemma 2 in the Appendix we will show that $E[\tau^2] < \infty$ for any initial condition for the TO policy when $E[X^2] < \infty$.

Thus we obtain

$$Sup_{(0 \leq \alpha < 1)} v_\alpha(q,e) - v_\alpha(0,\bar{e}) < \infty$$

for each $(q,e)$. This proves (9).

Now we show that $v_\alpha(q,e)$ is non-decreasing in $q$ and non-increasing in $e$. Let $v_n$ be $n$-step $\alpha$-discount optimal cost where $v_0 = c$, a constant. Then $v_n$ satisfies

$$v_{n+1}(q,e) = min_t\{q + \alpha E[v_n((q - g(t))^+ + X, e - t + Y)]\}. \tag{10}$$

To prove our assertion, we use induction. $v_0(q,e)$ satisfies the required properties. Let $v_n(q,e)$ also does. Then from (10) it is easy to show that $v_{n+1}(q,e)$ also satisfies these monotonicity properties. We have shown above that

$$v_\alpha(q,e) = \lim_{n \to \infty} v_n(q,e).$$

Thus $v_\alpha(q,e)$ inherits these properties. ∎

In Section III we identified a throughput optimal policy when $g$ is nondecreasing, concave. Theorem

3 guarantees the existence of an optimal mean delay policy. It is of interest to identify one such policy also. In general one can compute an optimal policy numerically via Value Iteration or Policy Iteration but that can be computationally intensive (especially for large data and energy buffer sizes). Also it does not provide any insight and requires traffic and energy profile statistics. In Section III we also provided a greedy policy (6) which is very intuitive, and is throughput optimal for linear $g$. However for concave $g$ (including the cost function $\frac{1}{2}log(1 + \gamma t)$) it is *not* throughput optimal and provides low mean delays only for low load. Next we show that it provides minimum mean delay for linear $g$.

**Theorem 4** The Greedy policy (6) is $\alpha$-discount optimal for $0 < \alpha < 1$ when $g(t) = \gamma t$ for some $\gamma > 0$. It is also average cost optimal.

**Proof:** We first prove the optimality for $0 < \alpha < 1$ where the cost function is

$$J_\alpha(\delta, q, e) = E\left[\sum_{k=0}^{\infty} \alpha^k q_k\right]$$

for a policy $\delta$. Let there be an optimal policy that violates (6) at some time $k$, i.e., $t_k \neq min\ (\frac{q_k}{\gamma},\ E_k)$. Clearly $t_k \leq E_k$. Also taking $t_k > q_k/\gamma$ wastes energy and hence cannot be optimal. The only possibility for an optimal policy to violate (6) is when $t_k < q_k/\gamma$ and $q_k/\gamma \leq E_k$. This is done with the hope that using the extra energy $\tilde{t}_k - t_k$ (where $\tilde{t}_k \overset{\triangle}{=} q_k/\gamma$) later can possibly reduce the cost. However this *increases* the total cost by at least

$$\gamma \alpha^k(\tilde{t}_k - t_k) - \gamma \alpha^{k+1}(\tilde{t}_k - t_k) = \gamma \alpha^k(\tilde{t}_k - t_k)(1 - \alpha) > 0$$

on that sample path. Thus such a policy can be improved by taking $t_k = \tilde{t}_k$. This holds for any $\alpha$ with $0 < \alpha < 1$. Also, from Theorem 3, under the conditions given there, an $\alpha$-discount optimal policy converges to an average cost optimal policy as $\alpha \nearrow 1$. This shows that (6) is also average cost optimal. ∎

The fact that Greedy is $\alpha$-discount optimal as well as average cost optimal implies that it is good not only for long term average delay but also for transient mean delays.

## V. GENERALIZATIONS

In this section we consider two generalizations. First we will extend the results to the case of fading channels and then to the case where the sensing and the processing energy at a sensor node are non-negligible with respect to the transmission energy.

In case of fading channels, we assume flat fading during a slot. In slot $k$ the channel gain is $h_k$. The sequence $\{h_k\}$ is assumed stationary, ergodic, independent of the traffic sequence $\{X_k\}$ and the energy generation sequence $\{Y_k\}$. Then if $T_k$ energy is spent in transmission in slot $k$, the $\{q_k\}$ process evolves as

$$q_{k+1} = (q_k - g(h_k T_k))^+ + X_k.$$

If the channel state information (CSI) is not known to the sensor node, then $T_k$ will depend only on $(q_k, E_k)$. One can then consider the policies used above. For example we could use $T_k = min(E_k, E[Y] - \epsilon)$. Then the data queue is stable if $E[X] < E[g(h(E[Y] - \epsilon))]$. We will call this policy unfaded TO. If we use Greedy (6), then the data queue is stable if $E[X] < E[g(hY)]$.

If CSI $h_k$ is available to the node at time $k$, then the following are the throughput optimal policies. If $g$ is linear, then $g(x) = \beta x$ for some $\beta > 0$. Then, if $0 \leq h \leq \bar{h} < \infty$ and $P(h = \bar{h}) > 0$, the optimal policy is: $T(\bar{h}) = (E[Y] - \epsilon)/p(h = \bar{h})$ and $T(h) = 0$ otherwise. Thus if $h$ can take an arbitrarily large value with positive probability, then $E[hT(h)] = \infty$ at the optimal solution.

If $g(x) = \frac{1}{2} log(1 + \beta x)$, then the water filling (WF) policy

$$T_k(h) = \left( \frac{1}{h_0} - \frac{1}{h} \right)^+ \tag{11}$$

with the average power constraint $E[T_k] = E[Y] - \epsilon$, is throughput optimal because it maximizes $\frac{1}{2} E_h[log(1 + \beta h T(h))]$ with the given constraints.

Both of the above policies can be improved as before, by not wasting energy when there is not enough data. As in (7) in Section III, we can further improve WF by taking

$$T_k = min \left( g^{-1}(q_k), E_k, \left( \frac{1}{h_0} - \frac{1}{h} + 0.001(E_k - cq_k)^+ \right)^+ \right). \tag{12}$$

We will call it MWF. These policies will not minimize mean delay. For that, we can use the MDP framework used in Section IV and numerically compute the optimal policies.

Till now we assumed that all the energy that a node consumes is for transmission. However, sensing, processing and receiving (from other nodes) also require significant energy, especially in more recent higher end sensor nodes ([23]). Since we have been considering a single node so far, we will now include the energy consumed by sensing and processing only. For simplicity, we will assume that the node is always in one energy mode (e.g., lower energy modes [28] available for sensor nodes will not be considered). If a sensor node with an energy harvesting system can be operated in energy neutral operation in normal mode itself (i.e., it satisfies the conditions in Lemma 1), then there is no need to have lower energy modes. Otherwise one has to resort to energy saving modes.

We will assume that $Z_k$ is the energy consumed by the node for sensing and processing in slot $k$. Unlike $T_k$ (which can vary according to $q_k$), $\{Z_k\}$ can be considered a stationary ergodic sequence. The rest of the system is as in Section II. Now we briefly describe a energy management policy which is an extension of the TO policy in Section III. This can provide an energy neutral operation in the present case. Improved/optimal policies can be obtained for this system also but will not be discussed due to lack of space.

Let $c$ be the minimum positive constant such that $E[X] < g(c)$. Then if $c + E[Z] < E[Y] - \delta$, (where $\delta$ is a small positive constant) the system can be operated in energy neutral operation: If we take $T_k \equiv c$ (which can be done with high probability for all $k$ large enough), the process $\{q_k\}$ will have a unique stationary, ergodic distribution and there will always be energy $Z_k$ for sensing and processing for all $k$ large enough. The result holds if $\{(X_k, Y_k, Z_k)\}$ is an ergodic stationary sequence. The arguments to show this are similar to those in Section III and are omitted.

When the channel has fading, we need $E[X] < E[g(ch)]$ in the above paragraph.

## VI. Simulations

In this section, we compare the different policies we have studied via simulations. The $g$ function is taken as linear $(g(x) = 10x)$ or as $g(x) = log(1 + x)$ . The sequences $\{X_k\}$ and $\{Y_k\}$ are iid. (We have also done limited simulations when $\{X_k\}$ and $\{Y_k\}$ are Autoregressive and found that conclusions drawn in this section continue to hold). We consider the cases when $X$ and $Y$ can have exponential, uniform, Erlang or Hyperexponential distributions. The policies considered are: Greedy, TO, $T_k \equiv Y_k$, MTO (with $c = 0.1$) and the mean delay optimal. At the end, we will also consider channels with fading. For the linear $g$, we already know that the Greedy policy is throughput optimal as well as mean delay optimal.

The mean queue lengths for the different cases are plotted in Figs. 2-10.

In Fig. 2, we compare Greedy, TO and mean-delay optimal (OP) policies for nonlinear $g$. The OP was computed via Policy Iteration. For numerical computations, all quantities need to be finite. So we took data and energy buffer sizes to be $50$ and used quantized versions of $q_k$ and $E_k$. The distribution of $X$ and $Y$ is Poisson truncated at $5$. These changes were made only for this example. Now $g(E[Y]) = 1$ and $E[g(Y)] = 0.92$. We see that the mean queue length of the three policies are negligible till $E[X] = 0.8$.

After that, the mean queue length of the Greedy policy rapidly increases while performances of the other two policies are comparable till 1 (although from $E[X] = 0.6$ till close to 1, mean queue length of TO is approximately double of OP). At low loads, Greedy has less mean queue length than TO.

Fig. 3 considers the case when $X$ and $Y$ are exponential and $g$ is linear. Now $E[Y] = 1$ and $g(E[Y]) = E[g(Y)] = 10$. Now all the policies considered are throughput optimal but their delay performances *differ*. We observe that the policy $T_k \equiv Y_k$ (henceforth called unbuffered) has the worst performance. Next is the TO.

Fig. 4 plots the case when $g$ is linear and $X$ and $Y$ are uniformly distributed. $E[Y] = 1$ and $g(E[Y]) = E[g(Y)] = 10$. Although the comparative performance of the four policies is as in Fig. 3, performances of the three policies are somewhat closer for this case. An interesting observation is that although the mean delay of the Greedy for exponential distribution is close to that of the uniform case, for the unbuffered and the TO policies, the mean delay of the exponential is much worse.

Figs. 5 and 6 provide the above results for $g$ nonlinear. When $X$ and $Y$ are exponential, the results are provided in Fig. 5 and when they are Erlang (obtained by summing 5 exponentials), they are in Fig. 6. Now, as before $T_k \equiv Y_k$ is the worst. The Greedy performs better than the other policies for low values of $E[X]$. But Greedy becomes unstable at $E[g(Y)]$ (= 2.01 for Fig. 5 and = 2.32 for Fig. 6) while the throughput optimal policies become unstable at $g(E[Y])$ (= 2.40 for Fig. 5 and Fig. 6). Now for higher values of $E[X]$, the modified TO performs the best and is close to Greedy at low $E[X]$.

Figs. 7-10 provide results for fading channels. The fading process $\{h_k\}$ is iid taking values $0.1, 0.5, 1.0$ and $2.2$ with probabilities $0.1, 0.3, 0.4$ and $0.2$ respectively. Figs. 7, 8 are for the linear $g$ and Figs. 9, 10 are for the nonlinear $g$. The policies compared are unbuffered, Greedy, Unfaded TO (6) and Fading TO (WF) (11) . In Figs. 9 and 10, we have also considered Modified Unfaded TO (7) and Modified Fading TO (MWF) (12).

In Fig. 7, $X$ and $Y$ are Erlang distributed. For this case, $E[Y] = 1$, $E[g(hY)] = 10$ and $E[g(hE[Y])] = 10$. We see that the stability region of fading TO is $E[X] < E[g(\bar{h}Y)]$ (= 22.0) while that of the other three algorithms is $E[X] < 10$. However, mean queue length of fading TO is also larger from the beginning till almost 10. This is because in fading TO, we transmit only when $h = \bar{h} = 2.2$ which has a small probability (= 0.2) of occurence.

In Fig. 8, $X$ and $Y$ have Hyperexponential distributions. The distribution of r.v. $X$ is a mixture of 5 exponential distributions with means $E[X]/4.9, 2E[X]/4.9, 3E[X]/4.9, 6E[X]/4.9$ and $10E[X]/4.9$ and probabilities $0.1, 0.2, 0.2, 0.3$ and $0.2$ respectively. The distribution of $Y$ is obtained in the same way. Now $E[Y] = 1$, $E[g(hY)] = 10$ and $E[g(hE[Y])] = 10$. We observe the same trends here as in Fig. 7 except that the mean queue lengths of the different algorithms vary much more in Fig. 8 when compared to Fig. 7. Also, except for Fading TO the mean queue lengths in Fig. 8 are much more than in Fig. 7. This is expected because the Hyperexponential distribution has much more variability than Erlang.

Figs. 9 and 10 consider nonlinear $g$. In Fig. 9 $X, Y$ are Erlang distributed and in Fig. 10 $X, Y$ are Hyperexponential as in Figs. 7 and 8. In Fig. 9, $E[Y] = 1, E[g(hY)] = 0.62, E[g(hE[Y])] = 0.64$ while in Fig.10, $E[Y] = 1, E[g(hY)] = 0.51$ and $E[g(hE[Y])] = 0.64$. Now we see that the stability region of unbuffered and Greedy is the smallest, then of TO and MTO while WF and MWF provide the largest region and are stable for $E[X] < 0.70$. MTO and MWF provide improvements in mean queue lengths over TO and WF. The difference in stability regions is smaller for Erlang distribution.

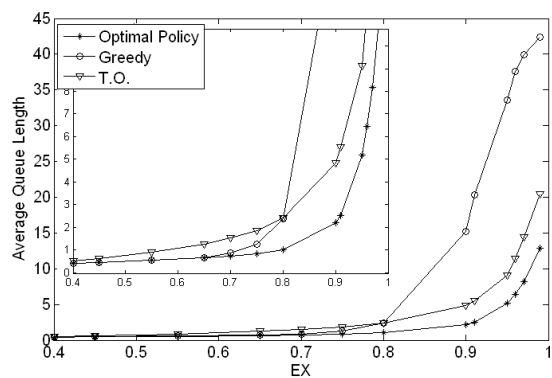

Fig. 2. Mean Delay Optimal, Greedy, TO Policies with No Fading; Nonlinear $g$; Finite, Quantized data and energy buffers; $X, Y$: Poisson truncated at 5; $E[Y] = 1, E[g(Y)] = 0.92, g(E[Y]) = 1$

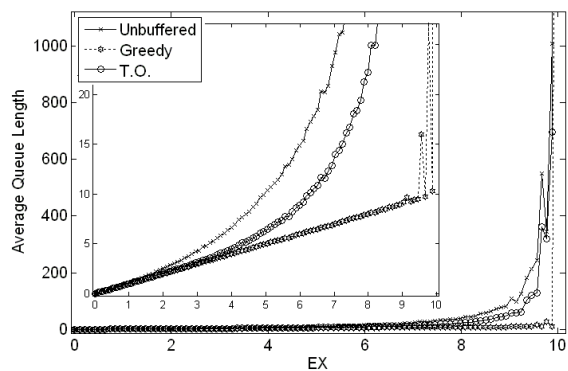

Fig. 3. Comparison of policies with No Fading; $g(x) = 10x$; $X, Y$: Exponential; $E[Y] = 1, E[g(Y)] = 10, g(E[Y]) = 10$

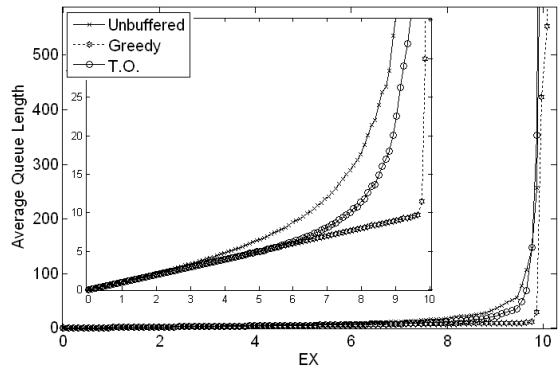

Fig. 4. Comparison of policies with No Fading; $g(x) = 10x$; $X, Y$: Uniform; $E[Y] = 1, E[g(Y)] = 10, g(E[Y]) = 10$

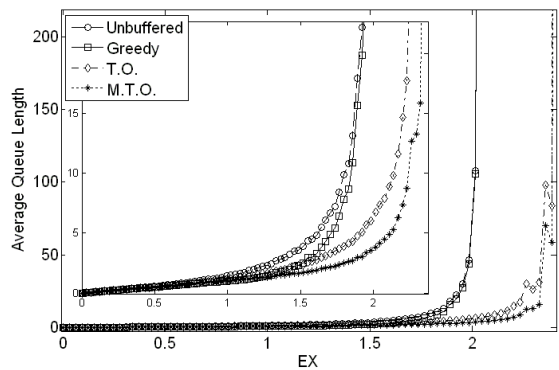

Fig. 5. Comparison of policies with No Fading; $g(x) = log(1 + x)$; $X, Y$: Exponential; $E[Y] = 10, E[g(Y)] = 2.01, g(E[Y]) = 2.4$

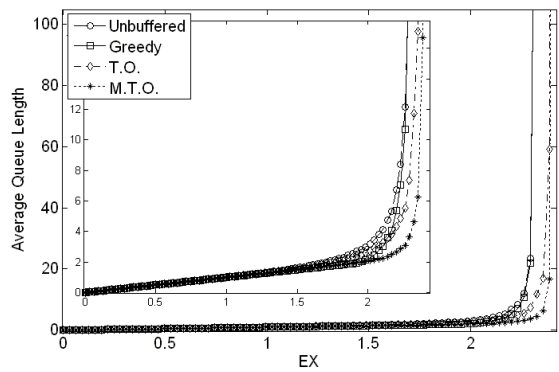

Fig. 6. Comparison of policies with No Fading; $g(x) = log(1 + x)$; $X, Y$: Erlang(5); $E[Y] = 10, E[g(Y)] = 2.32, g(E[Y]) = 2.4$

## VII. CONCLUSIONS

We have considered a sensor node with an energy harvesting source, deployed for random field estimation. Throughput optimal and mean delay optimal energy management policies are identified which can make the system work in energy neutral operation. The mean delays of these policies are compared

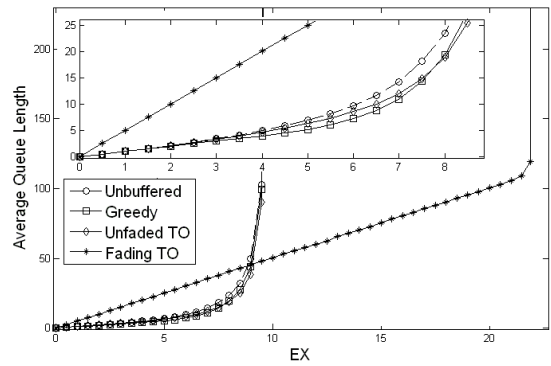

Fig. 7.   Comparison of policies with Fading; $g(x) = 10x$; $X, Y$: Erlang(5); $E[Y] = 1, E[g(Y)] = 10, g(E[Y]) = 10$

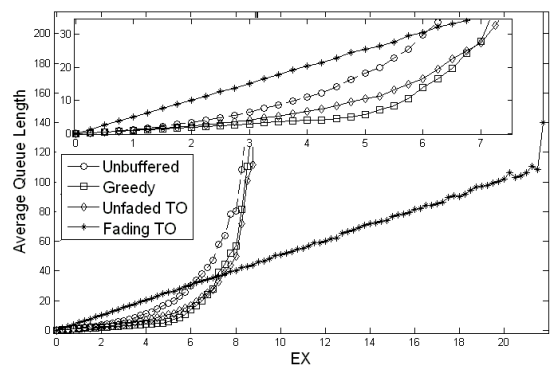

Fig. 8.   Comparison of policies with Fading; $g(x) = 10x$; $X, Y$: Hyperexponential(5); $E[Y] = 1, E[g(Y)] = 10, g(E[Y]) = 10$

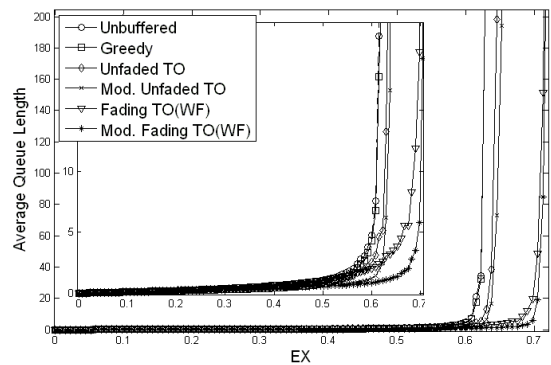

Fig. 9.   Comparison of policies with Fading; $g(x) = log(1 + x)$; $X, Y$: Erlang(5); $E[Y] = 1, E[g(hY)] = 0.62, E[g(hE[Y])] = 0.64$; WF, Mod. WF stable for $E[X] < 0.70$

with other suboptimal policies via simulations. It is found that having energy storage allows larger stability region as well as lower mean delays.

We have extended our results to fading channels and when energy at the sensor node is also consumed in sensing and data processing. Similarly we can include leakage/wastage of energy when it is stored in the energy buffer and when it is extracted. Suitable MACs for such sensor nodes have also been studied in [27].

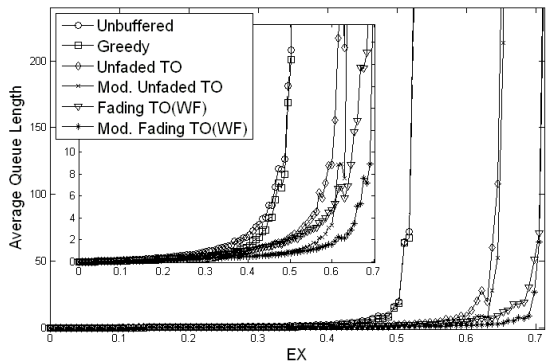

Fig. 10. Comparison of policies with Fading; $g(x) = log(1+x)$; $X, Y$: Hyperexponential(5); $E[Y] = 1, E[g(hY)] = 0.51, E[g(hE[Y])] = 0.64$; WF, Mod. WF stable for $E[X] < 0.70$

## VIII. Acknowledgement

This work was partially supported by a research grant from Boeing Corporation.

## IX. Appendix

To avoid trivialities we assume $P[X_k > 0] > 0$. For the following lemma we also assume that $P[X_k = 0] > 0$.

**Lemma 2** When $\{X_k\}, \{Y_k\}$ are iid, $E[X] < g(E[Y] - \epsilon), e \leq \bar{e}$, and $E[X^\alpha] < \infty$ for some $\alpha \geq 1$ then

$$\tau \triangleq inf\{k \geq 1 : (q_k, E_k) = (0, \bar{e})\}$$

satisfies $E[\tau^\alpha] < \infty$ for any $(q_0, E_0) = (q, e)$.

**Proof:** Let

$$A = \{(q, e) \ : \ q + e \leq \beta\}$$

where $\beta$ is an appropriately defined positive, finite constant. We will first show that starting from any initial $(q_0, E_0) = (q, e)$ the first time $\bar{\tau}$ to reach $A$ satisfies $E[\bar{\tau}^\alpha] < \infty$. Next we will show that with a positive probability in a finite (bounded) number of steps $(q_k, E_k)$ can reach from $A$ to $(0, \bar{e})$. Then by a standard coin tosing argument, we will obtain $E[\tau^\alpha] < \infty$.

To show $E[\bar{\tau}^\alpha] < \infty$, we use a result in [[13], pp.116]. Then it is sufficient to show that for $h(q, e) = q$,

$$sup_{(q,e) \notin A} E\left[h(q_1, E_1) - h(q, e)|q_0 = q, E_0 = e\right] < -\delta \qquad (13)$$

for some $\delta > 0$ and

$$E\left[|h(q_1, E_1) - h(q, e)|^\alpha|(q_0, E_0) = (q, e)\right] < \infty \qquad (14)$$

for all $(q, e)$.

Instead of using (13), (14) on the Markov chain $\{(q_k, E_k)\}$ we use it on the Markov chain $\{(q_{Mk}, E_{Mk}), k \geq 0\}$ where $M > 0$ is an appropriately large postive integer. Thus for (14) we have to show that

$$E[|q_M - q|^\alpha|q_0 = q] < \infty$$

which holds if $E[X^\alpha] < \infty$.

Next we show (13). Taking $\beta$ large enough, since $T_k \leq \bar{e}$, we get for $(q, e) \notin A$,

$$E\left[h(q_M, E_M) - h(q_0, E_0)|(q_0, E_0) = (q, e)\right]$$
$$= E\left[q + \sum_{n=0}^{M}(X_n - g(T_n)) - q|(q_0, E_0) = (q, e)\right].$$

Thus, (13) is satisfied if

$$E[X_1] < \frac{1}{M}\sum_{k=1}^{M} E\left[g(T_n)|(q_0, E_0) = (q, e)\right] - \delta. \qquad (15)$$

But for TO,

$$\frac{1}{M}\sum_{k=1}^{M} E[g(T_n)|(q_0, E_0) = (q, e)]$$
$$= \frac{1}{M}\sum_{k=1}^{M} E[g(T_n)|E_0 = e] \rightarrow g(E[Y] - \epsilon)$$

and thus there is an $M$ (choosing one corresponding to $e = 0$ will be sufficient for other $e$) such that if $E[X] < g(E[Y] - \epsilon)$, then (15) will be satisfied for some $\delta > 0$.

Now we show that from any point $(q, e) \in A$, the process can reach the state $(0, \bar{e})$ with a positive probability in a finite number of steps. Choose positive $\epsilon_1, \epsilon_2, \epsilon_3, \epsilon_4$ such that $P[X_k = 0] = \epsilon_1 > 0$ and $P[Y_k > \epsilon_3] > \epsilon_4$, $g(\epsilon_3) = \epsilon_2$, where such positive constants exist under our assumptions. Then with probability $\geq (\epsilon_1 \epsilon_4)^{\left( \left[ \frac{\beta}{\epsilon_2} \right] + \left[ \frac{\bar{e}}{\epsilon_3} \right] \right)}, (q_k, E_k)$ reaches $(0, \bar{e})$ in $\left[ \frac{\beta}{\epsilon_2} \right] + \left[ \frac{\bar{e}}{\epsilon_3} \right]$ steps where $[x]$ denotes the smallest integer $\geq$ $x$. ∎

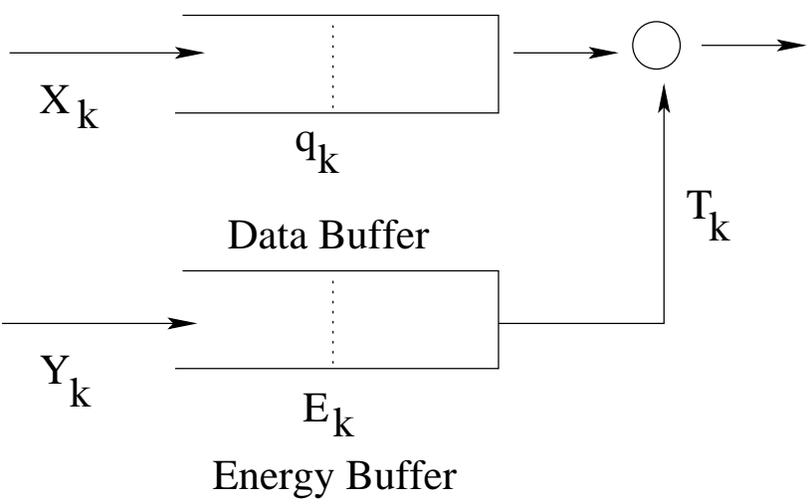